\documentclass[%
12pt,
reprint,
superscriptaddress,
preprintnumbers,
amsmath,amssymb,
aps,
prb,
floatfix,
]{revtex4-2}

\usepackage{txfonts}

\usepackage{graphicx}% Include figure files
\usepackage{rotating}
\usepackage{dcolumn}% Align table columns on the decimal point
\usepackage{bm}% bold math
\usepackage[colorlinks, allcolors=blue]{hyperref}% add hypertext capabilities
\usepackage[mathlines]{lineno}% Enable numbering of text and display math
\usepackage{orcidlink}
\usepackage{sidecap}

\usepackage{xcolor}
\usepackage{microtype}
\usepackage{graphicx}

\begin{document}

\title{
Elasticity and acoustic velocities of $\delta$-AlOOH\\
at extreme conditions: a methodology assessment
}

\author{Chenxing Luo\,\orcidlink{0000-0003-4116-6851}}
\affiliation{Department of Applied Physics and Applied Mathematics, Columbia University, New York, New York 10027, USA}

\author{Yang Sun\,\orcidlink{0000-0002-4344-2920}}
\email[]{yangsun@xmu.edu.cn}
\affiliation{Department of Physics, Xiamen University, Xiamen, 361005, China}

\author{Renata M.\ Wentzcovitch\,\orcidlink{0000-0001-5663-9426}}
\email[]{rmw2150@columbia.edu}
\affiliation{Department of Applied Physics and Applied Mathematics, Columbia University, New York, New York 10027, USA}
\affiliation{Department of Earth and Environmental Sciences, Columbia University, New York, New York 10027, USA}
\affiliation{Lamont--Doherty Earth Observatory, Columbia University, Palisades, New York 10964, USA}
\affiliation{Data Science Institute, Columbia University, New York, New York 10027, USA}
\affiliation{Center for Computational Quantum Physics, Flatiron Institute, New York, NY 10010, USA}

\date{\today}

\begin{abstract}
Hydrous phases play a fundamental role in the deep-water cycle on Earth. Understanding their stability and thermoelastic properties is essential for constraining their abundance using seismic tomography. However, determining their elastic properties at extreme conditions is notoriously challenging. The challenges stem from the complex behavior of hydrogen bonds under high pressures and temperatures ($P,T$s). In this study, we evaluate how advanced molecular dynamics simulation techniques can address these challenges by investigating the adiabatic elasticity and acoustic velocities of $\delta$-AlOOH, a critical and prototypical high-pressure hydrous phase. We compared the performances of three methods to assess their viability and accuracy. The thermoelastic tensor was computed up to 140~GPa and temperatures up to 2,700~K using molecular dynamics with a DeePMD machine-learning interatomic potential based on the SCAN meta-GGA functional. The excellent agreement with ambient condition single-crystal ultrasound measurements and the correct description of velocity changes induced by H-bond disorder-symmetrization transition observed at 10~GPa in Brillouin scattering measurements underscores the accuracy and efficacy of our approach.
\end{abstract}

\maketitle

\section{Introduction}

Water is an active participant in dynamic processes that shape the Earth. Typically incorporated in hydrous minerals or nominally anhydrous minerals (NAMs), water contributes to slab subduction and mantle convection by facilitating rock deformation and reducing melting temperatures of mantle silicates \citep{ohtaniRoleWaterEarth2020}. 
However, accurately quantifying water content in the mantle, particularly in the lower mantle, remains challenging due to the need for more relevant thermoelastic and sound velocity data for hydrous minerals necessary for interpreting seismic observations \citep{hackerSubductionFactoryTheoretical2003}.
The complexity of measurements under extreme pressures and temperatures ($P,T$s) is exacerbated by hydrogen bonds' intricate behavior, e.g., strong anharmonicity, disorder-symmetrization transition, transition to a proton superionic behavior, dehydration, all difficult to quantify and might affect mechanical properties significantly.

To evaluate a)~the challenges that arise in determining the thermoelastic properties and sound velocities in hydrous systems and b)~how advanced computational simulations may address them, we investigate $\delta$-AlOOH ($\delta$) \citep{suzukiNewHydrousPhase2000, ohtaniStabilityFieldNew2001}, a prototypical high-pressure hydrous mineral. According to previous studies, $\delta$-AlOOH withstands the extreme pressure of the deep-lower mantle (up to 140~GPa) \citep{ohtaniStabilityFieldNew2001, sanoAluminousHydrousMineral2008, duanPhaseStabilityThermal2018, pietDehydrationDAlOOHEarth2020, luoProbingStateHydrogen2024}.
Its Al-O octahedron forms a post-stishovite-like framework structure, representative of coexisting hydrous systems, e.g., MgSiO$_4$H$_2$ phase H \citep{tsuchiyaFirstPrinciplesPrediction2013}, $\epsilon$-FeOOH \citep{zhuHydrogenBondSymmetrizationBreakdown2017a}, and hydrous aluminous silica \citep{lakshtanovPoststishovitePhaseTransition2007, umemotoPoststishoviteTransitionHydrous2016, ishiiSuperhydrousAluminousSilica2022}.
It is the simplest and one of the most well-studied high-pressure hydrous minerals \citep{duanPhaseStabilityThermal2018, pietDehydrationDAlOOHEarth2020}. $\delta$ undergoes H-bond disorder-symmetrization transition at $\sim$5--20~GPa at 300~K \citep{sano-furukawaDirectObservationSymmetrization2018, luoInitioInvestigationHbond2022, luoProbingStateHydrogen2024} and high-temperature proton diffusion above 1,500~K \citep{luoProbingStateHydrogen2024}. % {\color{red}
This comprehensive understanding of $\delta$-AlOOH's stability, structural transitions, and unique challenges make it ideal for understanding how well advanced atomistic simulations can address elasticity and sound velocities for H-bonded systems under the mantle's extreme $P,T$s.

Previous studies have determined $\delta$'s elasticity and acoustic velocities in a limited $P$-$T$ regime, revealing velocity profiles similar to those of anhydrous minerals (e.g., perovskite-type MgSiO$_3$, stishovite-type SiO$_2$, etc.) rather than typical hydrous minerals or NAMs \citep{duanPhaseStabilityThermal2018, mashinoSoundVelocitiesDAlOOH2016}. However, they have anomalies, especially across the H-bond disorder-symmetrization transition \citep{tsuchiyaElasticPropertiesDAlOOH2009a, mashinoSoundVelocitiesDAlOOH2016}. Static \textit{ab initio} calculations based on density functional theory (DFT) with the generalized gradient approximation (GGA/PBE) \citep{liElasticOpticalProperties2006, tsuchiyaElasticPropertiesDAlOOH2009a} up to 60~GPa predicted an anomalous jump in the elasticity and acoustic velocities due to ``H-bond symmetrization'' at 30~GPa.
Brillouin scattering measurements of shear wave ($V_S$) and compressional wave ($V_P$) speeds on polycrystalline samples \citep{mashinoSoundVelocitiesDAlOOH2016} to 120~GPa and 36.3~GPa, respectively, confirm this overall trend. However, the pressure-induced anomalies were observed at a much lower pressure, $\sim$10~GPa. Meanwhile, a recent ultrasound measurement on single-crystal at ambient conditions \citep{wangSinglecrystalElasticityPhase2022} obtained velocities lower than the previous Brioullion scattering measurements. To reconcile these discrepancies and to determine these parameters under the lower mantle's more extreme $P,T$s, more accurate determinations of $\delta$-AlOOH's elastic and acoustic properties over a broader pressure and temperature range are necessary.

\textit{Ab initio} calculations using the quasiharmonic approximation and stress-strain relations have accurately predicted elastic properties of minerals at mantle conditions as long as harmonic vibrations are warranted at high pressures \citep{karkiFirstPrinciplesDeterminationElastic1999, wentzcovitchThermoelasticPropertiesMgSiO3Perovskite2004, wuQuasiharmonicThermalElasticity2011, luoCijPythonCode2021}. This method is not applicable for hydrous phases at mantle conditions, and one must resort to molecular dynamics (MD) simulations to address the strongly anharmonic behavior of H-bonds. As an active research topic, the effectiveness of several MD simulation
methods to compute high-temperature elasticity, including the strain vs.\ stress method, strain fluctuation under isobaric conditions \cite{sprikSecondorderElasticConstants1984, materzaniniSolidsThatAre2023}, and stress fluctuation under isentropic and isothermal conditions \citep{rayMolecularDynamicsCalculation1985, rayCalculationElasticConstants1986}, are subjects of ongoing debate.
\citet{clavierComputationElasticConstants2017} presented an insightful overview of these methodologies.
Additional discrepancies in finite-temperature elasticity may originate from definitions of elasticity under finite $P,T$s.
For instance, the difference between isothermal ($C^T_{ijkl}$) and adiabatic elastic tensors ($C^S_{ijkl}$) is a non-negligible distinction at temperatures exceeding a few thousand Kelvin; the difference between the thermodynamic elastic tensor ($A_{ijkl}$) and the effective elastic tensor ($C_{ijkl}$) \citep{luoInitioCalculationsThirdorder2022a} under finite pressure still needs clarification.
Given the need to derive sound velocities for meaningful comparisons with measurements, our objective here is also to derive the adiabatic effective elastic tensor, clarify these differences, and understand their implications.

In this study, we perform MD simulations using the Deep Potential (DP) neural network interatomic potential \citep{zhangDeepPotentialMolecular2018, zhangEndtoendSymmetryPreserving2018} developed \citep{luoProbingStateHydrogen2024} to accurately reproduce \textit{ab initio} forces and energies obtained with the strongly constrained and appropriately normed (SCAN) meta generalized gradient approximation (meta-GGA) functional \citep{furnessAccurateNumericallyEfficient2020}. We evaluate $\delta$'s thermoelastic properties and acoustic velocities over the broad $P,T$ range of the mantle, i.e., up to 140~GPa and 2,700~K.
SCAN-DPMD has proven effective in replicating measured pressure vs.\ volume ($P$-$V$) relations at high $P,T$s \citep{luoProbingStateHydrogen2024} for $\delta$-AlOOH, even with protons in the superionic regime.
In addition, our previous studies suggested that meta-GGA functionals can accurately predict ionic solids' elastic and mechanical properties at extreme pressure and temperatures, particularly for hydrous systems \citep{wangInitioStudyStability2024, wanThermoelasticPropertiesBridgmanite2024}.
Therefore, $\delta$'s thermoelastic properties should be equally well predicted.
Based on these results, we highlight different definitions of elasticity and clarify the implications of different approaches at high $P,T$s.

The paper is structured as follows: Section~\ref{sec:method} describes the methods used in this study, Section~\ref{sec:result} presents the studies' findings, and Section \ref{sec:conclusion} presents our conclusion.

\section{Method}
\label{sec:method}

Section~\ref{sec:formalisms} uses the full cartesian notation to describe the elastic tensor, while Section~\ref{sec:elastic-moduli-velocities} uses the Voigt notation to describe the aggregate (isotropic) elastic coefficients. These are the standard notations for discussing these topics.

\subsection{Formalisms for thermoelasticity}
\label{sec:formalisms}

Here, we summarize three major approaches for computing the \textit{adiabatic effective} thermoelastic tensor through MD simulations. While other methods exist, our study tests the techniques presented here.

\subsubsection{The stress-fluctuation formalism}
\label{sec:stress-fluctuation}

The \textit{thermodynamic} elastic tensor, $A_{ijkl}$ ($i,j,k,l = 1,2,3$) is defined as the second-order strain derivative of the free energy ($F$) density for a system in equilibrium in MD simulations with a fixed simulation box of volume $V$. It is given by \cite{rayStatisticalEnsemblesMolecular1984, rayMolecularDynamicsCalculation1985, clavierComputationThermalElastic2023, zhenDeformationFluctuationHybrid2012}
\begin{equation}
\begin{split}
        A_{ijkl} \, (T,V) =&~\frac{1}{V} \frac{\partial^2 F}{\partial\epsilon_{ij} \partial\epsilon_{kl}} = \langle A^\mathrm{B}_{ijkl} \rangle - \frac{V}{k_\mathrm{B} T} \big[ \langle \sigma_{ij} \, \sigma_{kl} \rangle \\&\quad \hbox{} - \langle\sigma_{ij}\rangle\,\langle\sigma_{kl} \rangle \big] + \frac{N k_\mathrm{B} T}{V} \big( \delta_{il}\delta_{jk} + \delta_{ik}\delta_{jl} \big) \,,
        \label{eq:stress-fluctuation}
\end{split}
\end{equation}
where ``$\langle \cdot \rangle$" denotes the ensemble average over the simulation run time, $\epsilon_{ij}$ denotes the infinitesimal strain tensor, $N$ denotes the number of atoms, and $T$ denotes the temperature. The stress and Born matrix tensors ($\sigma_{ij}$ and $A^\mathrm{B}_{ijkl}$) are defined as the \textit{instantaneous} first- and second-order derivate to the internal energy density ($U/V$) w.r.t.\ infinitesimal strains $\epsilon_{ij}$, i.e.,
\begin{equation}
    \sigma_{ij} \equiv  \frac{1}{V}\frac{\partial U}{\partial\epsilon_{ij}}
    \quad\mathrm{and}\quad
    A^\mathrm{B}_{ijkl} \equiv \frac{1}{V}\frac{\partial^2 U}{\partial\epsilon_{ij}\partial\epsilon_{kl}}
    \, .
\end{equation}
They are evaluated and recorded periodically during an equilibrated MD run.
In our calculation, $A^\mathrm{B}_{ijkl}$ is recorded every 200~MD steps and $\sigma_{ij}$ is recorded every 20~steps.
We evaluate $A^\mathrm{B}_{ijkl}$ through numerical differentiation via \cite{clavierComputationThermalElastic2023}
\begin{equation}
     A^\mathrm{B}_{ijkl} = \frac{\partial\sigma_{ij}}{\partial\epsilon_{kl}} + \sigma_{il} \delta_{jk} + \sigma_{ik} \delta_{jl} \, .
\end{equation}
For a simulation run with an average stress $\sigma_{ij}$,
the \textit{effective} elastic tensor $C_{ijkl}$ is given by \cite{luoInitioCalculationsThirdorder2022a, dahlenTheoreticalGlobalSeismology1998, barronSecondorderElasticConstants1965},
\begin{equation}
\begin{split}
    C_{ijkl} = & \frac{1}{V} \frac{\partial^2 F}{\partial\epsilon_{ij} \partial\epsilon_{kl}} - \sigma_{ij} \delta_{kl} \\& \hbox{} + \tfrac{1}{2} (\sigma_{ik}\delta_{jl} + \sigma_{kj} \delta_{il} + \sigma_{il} \delta_{jk} + \sigma_{lj}\delta_{ik}) \,,
\end{split}
\end{equation}
it is further simplified when the stress is hydrostatic, i.e., $\sigma_{ij} = - P \, \delta_{ij}$ \citep{barronSecondorderElasticConstants1965},
\begin{equation}
    C_{ijkl} = \frac{1}{V} \frac{\partial^2 F}{\partial\epsilon_{ij} \partial\epsilon_{kl}} + P \, \big(\delta_{ij}\delta_{kl} - \delta_{il}\delta_{kj} - \delta_{ik}\delta_{jl} \big) \,,
    \label{eq:elastic-tensor}
\end{equation}
where $P$ denotes the pressure $P \equiv - \frac{1}{3} \mathrm{Tr}(\sigma_{ij})$.
$A_{ijkl}$ is not equal to $C_{ijkl}$ unless the system is free of any external pressure or stress.
For an $NVE$ run, Eqs.~\eqref{eq:stress-fluctuation} and \eqref{eq:elastic-tensor} give the adiabatic elastic tensors $A^S_{ijkl}$ and $C^S_{ijkl}$, and for an $NVT$ run \cite{rayCalculationElasticConstants1986}, Eqs.~\eqref{eq:stress-fluctuation} and \eqref{eq:elastic-tensor} give the isothermal elastic tensors $A^T_{ijkl}$ and $C^T_{ijkl}$ \cite{clavierComputationThermalElastic2023}.

\subsubsection{The stress vs.\ strain method}

Alternatively, the \textit{effective} thermoelastic tensor, $C_{ijkl}$, can be evaluated using the stress vs.\ strain approach. The components of $C_{ijkl}$ are obtained via numerical differentiation of the recorded stress tensor ($\sigma_{ij}$) with respect to the strain tensor ($\epsilon_{kl}$), i.e.,
\begin{equation}
    C_{ijkl} = \partial\sigma_{ij} / \partial\epsilon_{kl} \,,
    \label{eq:stress-strain}
\end{equation}
where $\sigma_{ij}$ denotes the stress tensor components, and $\epsilon_{kl}$ denotes the imposed strain tensor.
For an $NVE$ run, Eq.~\eqref{eq:stress-strain} gives the adiabatic elastic tensor $C^S_{ijkl}$, and for an $NVT$ run \cite{rayCalculationElasticConstants1986}, Eq.~\eqref{eq:stress-strain} gives the isothermal elastic tensor and $C^T_{ijkl}$.

\subsubsection{Adiabatic correction}

If one has already derived the isothermal tensor, the adiabatic tensor could be derived by applying the adiabatic correction.
For an orthorhombic system, the diagonal and off-diagonal components of the adiabatic tensor, $C_{ijkl}^S$, are connected to the isothermal tensor, $C_{ijkl}^T$, by \citep{wuQuasiharmonicThermalElasticity2011, daviesEffectiveElasticModuli1974, luoCijPythonCode2021}
\begin{equation}
    C_{iikk}^S = C_{iikk}^T + \frac{TV}{ C_V} \, \lambda_{i} \, \lambda_{k} \quad (i,k = 1,2,3) \,.
    \label{eq:adiabatic-correction}
\end{equation}
Here, $\lambda_{i}$ is given by \citep{daviesEffectiveElasticModuli1974}
\begin{equation}
    \lambda_{i} = -\alpha_{k} \, C_{iikk}^T \,,
\end{equation}
where $\alpha_k$, the linear thermal expansion coefficient at constant pressure, is defined by
\begin{equation}
    \alpha_i =  \frac{\partial \ln a_i}{\partial T} \, \bigg \vert_P\,
\end{equation}
and is obtained by numerical differentiation of the lattice parameter, $a(T)$, w.r.t.\ $T$.
$C_V$ denotes the constant volume-specific heat. It can be calculated in the $NVT$ ensemble using \citep{klochkoGeneralRelationsObtain2021}
\begin{equation}
    N C_V = \frac{\mathrm{var}(U)}{k_\mathrm{B} T^2} = \frac{1}{k_\mathrm{B} T^2} \left[\big\langle U^2 \big\rangle - \big\langle U \big\rangle^2 \right] \,,
    \label{eq:volume-specific-heat-capacity}
\end{equation}
where $N$ denotes the number of atoms and $U$ denotes the internal energy. This correction is temperature-dependent and is more relevant at high temperatures.

\subsection{Elastic moduli and acoustic velocities}
\label{sec:elastic-moduli-velocities}

The bulk modulus ($K$) and shear modulus ($G$) are derived from the $C_{ij}$ as Voigt-Reuss-Hill (VRH) averages \citep{hillElasticBehaviourCrystalline1952}, i.e.,
\begin{equation}
    K_\mathrm{H} = \tfrac{1}{2}(K_\mathrm{V} + K_\mathrm{R}) \quad \mathrm{and} \quad  G_\mathrm{H} = \tfrac{1}{2}(G_\mathrm{V} + G_\mathrm{R}) ~,
    \label{eq:hill-elastic-moduli}
\end{equation}
where $K_\mathrm{H}$, $G_\mathrm{H}$ represents the Hill averages; the Voigt averages, $K_\mathrm{V}$, $G_\mathrm{V}$, and Reuss averages $K_\mathrm{R}$, $G_\mathrm{R}$, are given by \citep{hillElasticBehaviourCrystalline1952, wattPOLYXSTALFORTRANProgram1987}
\begin{subequations}
\label{eq:voigt-reuss-elastic-moduli}
\begin{align}
9 \, K_\mathrm{V} &= (c_{11}+c_{22}+c_{33}) + 2(c_{12}+c_{23}+c_{31}) \,, \\
15 \, G_\mathrm{V} &=  (c_{11} + c_{22} + c_{33})  -   (c_{12} + c_{23} + c_{31}) + 3 (c_{44} + c_{55} + c_{66}) \,, \\
1 / K_\mathrm{R} &= (s_{11}+s_{22}+s_{33})+2(s_{12}+s_{23}+s_{31}) \,, \\
15 / G_\mathrm{R} &= 4 (s_{11} + s_{22} + s_{33}) - 4 (s_{12} + s_{23} + s_{31}) + 3 (s_{44} + s_{55} + s_{66}) \,, 
\end{align}
\end{subequations}
where $s_{ij}$ represents the components of the elastic compliance tensor, which are inversely related to the stiffness tensor ($S_{ij} = C_{ij}^{-1}$).
The acoustic wave speeds, $V_S$ and $V_P$, are
\begin{equation}
    V_P = \sqrt{(K + \tfrac{3}{4} G)/ \rho} \quad \mathrm{and} \quad  V_S = \sqrt{G / \rho} ~.
    \label{eq:velocity}
\end{equation}

\subsection{DPMD simulations}

DPMD simulations were performed in \textsc{LAMMPS} \citep{thompsonLAMMPSFlexibleSimulation2022} with Deep Potential (DP) neural network interatomic potential \citep{zhangDeepPotentialMolecular2018, zhangEndtoendSymmetryPreserving2018} implemented in the \textsc{DeePMD-kit} v2.0 package \citep{wangDeePMDkitDeepLearning2018, zengDeePMDkitV2Software2023}.
The potential was trained based on the SCAN meta-GGA functional's \citep{furnessAccurateNumericallyEfficient2020} description of $\delta$'s force and energy implemented in the \textsc{Vienna Ab-initio Simulation Package} (VASP) \citep{kresseEfficientIterativeSchemes1996} with 520~eV energy cutoff for plane-wave basis set sampled over a $2\times2\times2$ Monkhorst-Pack $k$-point mesh for 128-atom supercells \citep{luoProbingStateHydrogen2024}.
At 3,000~K, the highest benchmarked temperature used, this SCAN-DP reaches an accuracy of $\sim$2~meV/atom RMSE in energy and $\sim$0.12~eV/\AA$^3$ RMSE in force compared to SCAN-DFT \citep{luoProbingStateHydrogen2024}.

The components of the $C_{ij}$ tensor were determined in a dense $P,T$-mesh, every 5~GPa up to 150~GPa, and every 600~K from 300~K to 2,700~K using SCAN-DPMD on 8,192-atom supercells. To obtain the cell shape at a given temperature, constant pressure ($NPT$) DPMD simulations were performed for 0.1~ns with a timestep of 0.2~fs and the Nos\'e–Hoover thermo-baro-stat that incorporates modular invariance \citep{hooverKineticMomentsMethod1996, martynaConstantPressureMolecular1994, wentzcovitchInvariantMoleculardynamicsApproach1991} to equilibrate the cell shape at given $P,T$s. For subsequent simulations, we start from a cell shape equal to those equilibrated in the $NPT$ runs.

To derive the elastic tensor using the stress-fluctuation method [Eq.~\eqref{eq:stress-fluctuation} and \eqref{eq:elastic-tensor}], we performed $NVT$ simulations to generate atomic coordinates and velocities at desired temperatures. These runs were followed by $NVE$ runs for at least 4~ns with a timestep of 0.5~fs to determine the components of the \textit{adiabatic} $C^S_{ij}$ tensor under the specified conditions.

To derive the adiabatic (isothermal) elastic tensor using the stress vs.\ strain method [Eq.~\eqref{eq:stress-strain}], we started from a cell shape equilibrated with constant pressure MD at specific $P,T$ conditions. Then, fixed the cell shape and conducted $NVT$ simulations to reach the target equilibrium temperature.
After removing the thermostat, we deformed the cell shape by imposing a $1\%$ strain, followed the state with an $NVE$ ($NVT$) run, and recorded the average stress response over an MD run of 12.5~ps with a timestep of 0.5~fs.
Details regarding the calculation of $C_V$ for adiabatic correction [Eqs.~\eqref{eq:adiabatic-correction}--\eqref{eq:volume-specific-heat-capacity}] were presented earlier in our previous study \citep{luoProbingStateHydrogen2024}.

\section{Result and discussion}
\label{sec:result}

\subsection{Compressive behavior and thermal expansion}

\begin{figure}[htbp]
    \centering
    \includegraphics[width=.46\textwidth]{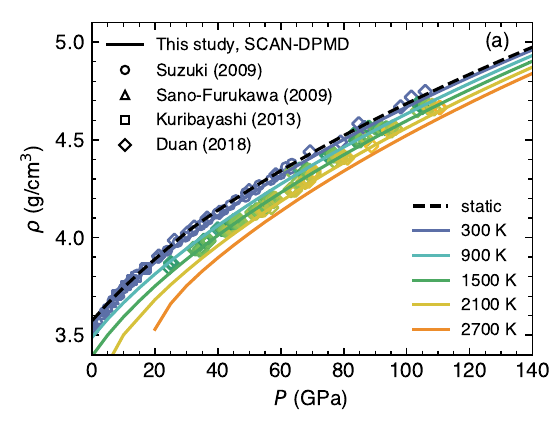}
    \includegraphics[width=.48\textwidth]{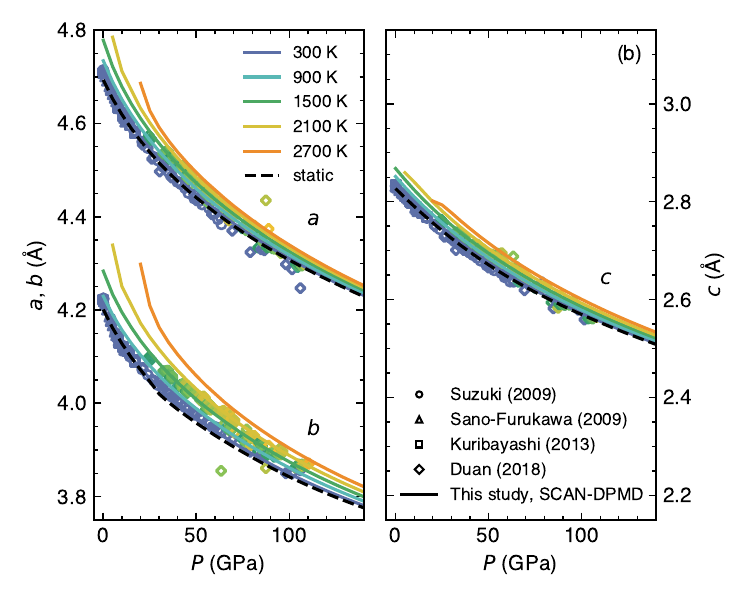}
    \caption{Compressive behavior of $\delta$-AlOOH at 300--2,700~K and under static condition predicted by SCAN-DPMD compared to X-ray diffraction measurements by \citet{sano-furukawaChangeCompressibilityDAlOOH2009, suzukiCompressibilityHighpressurePolymorph2009, kuribayashiObservationPressureinducedPhase2014} (300~K) and by \citet{duanPhaseStabilityThermal2018} (300--2,500~K): (a)~isothermal density-pressure relations ($\rho$ vs.\ $P$); (b)~isothermal axial compressive behavior ($a, b, c$ vs.\ $P$).}
    \label{fig:1}
\end{figure}

In Fig.~\ref{fig:1}, we evaluate the accuracy of SCAN-DPMD in predicting the compressive behavior of the system by comparing its predictions with X-ray diffraction measurements \cite{sano-furukawaChangeCompressibilityDAlOOH2009, suzukiCompressibilityHighpressurePolymorph2009, kuribayashiObservationPressureinducedPhase2014, duanPhaseStabilityThermal2018}, available over a wide $P,T$ range. Both density-pressure relations [$\rho$ vs.\ $P$, Fig.~\ref{fig:1}(a)] and axial compressions ($a, b, c$ vs.\ $P$, Fig.~\ref{fig:1}(b)) are accurately reproduced in DPMD simulations at finite temperatures.

$\delta$'s 300~K compression curve was measured extensively below 30~GPa. Our results agree well with these measurements. In particular, we fully reproduced both the compressibility change resulting from the $a$ and $b$ axes hardening around 10~GPa \cite{sano-furukawaChangeCompressibilityDAlOOH2009}, and the more subtle change in the $c$ axis compressibility. These compressibility changes are less noticeable at $900 < T < 1,500$~K.  However, anomalies reappear at $2,100 < T < 2,700$~K for pressures below 40~GPa.
For $P > 40$ GPa, no change in compressive behavior is seen at any $T$.
Above 60~GPa, the discrepancies between SCAN-DPMD predictions and experimental data slightly increase, likely due to the more significant uncertainties in the measurements, evident from the wider spread of experimental data at higher pressures \citep{duanPhaseStabilityThermal2018}. The differences between SCAN-DPMD prediction and measurements are comparable to experimental uncertainties. The increasing uncertainties in the axial compressive behavior measurements highlight experimental challenges at elevated pressures.

\begin{figure}[htbp]
    \centering
    \includegraphics[width=.48\textwidth]{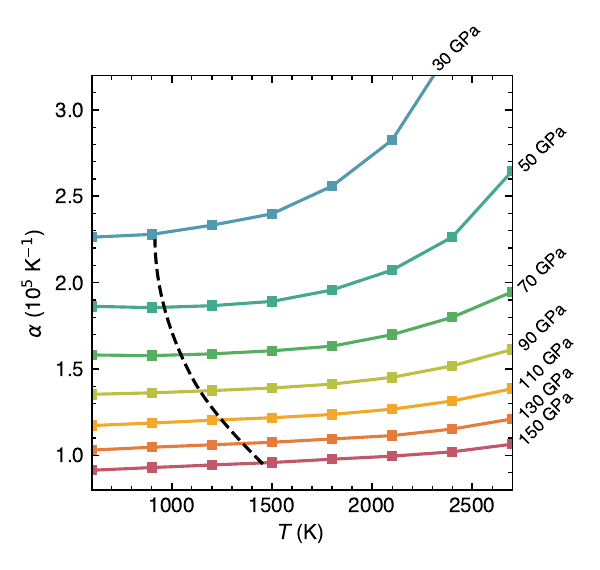}
    \caption{Thermal expansivity ($\alpha$) of $\delta$-AlOOH predicted by SCAN-DPMD. The dashed black curve indicates the critical temperatures where the curvature of $\alpha(T)$ alternates, defined by $\partial^2\alpha/\partial T^2\big|_P = 0$.}
    \label{fig:alpha-temp}
\end{figure}

Using these results, we determined $\delta$'s volume thermal expansivity,
$\alpha = \partial \ln V / \partial T \big|_P$,
via numerical differentiation (see Fig.~\ref{fig:alpha-temp}). Our calculations indicate $\delta$'s $\alpha$ decreases with pressure, from $\sim$$2.2\times10^{-5}$~K$^{-1}$ at 30~GPa to $\sim$$0.9\times10^{-5}$~K$^{-1}$ at 150~GPa. At high temperatures, first, $\delta$'s $\alpha$ increases gently. Above $\sim$1,000--1,500~K, and depending on the pressure, it abruptly increases with $T$, suggesting significant anharmonicity at high temperatures. These positive curvatures of $\alpha(T)$, i.e., $\partial^2\alpha/\partial T^2\big|_P > 0$, usually occur as a precursor to phase transitions, and in this case, correlated with the melting of the H sublattice that leads to the diffusive boundary to superionic behavior reported earlier \citep{luoProbingStateHydrogen2024}. Our calculation suggests that the curvature increases significantly at $\sim$1,000~K at 30~GPa and 1,500~K at 150~GPa. This temperature is lower than the normal mantle and the slab geotherm \citep{brownThermodynamicParametersEarth1981, eberleNumericalStudyInteraction2002}.

In summary, the consistency between our SCAN-DPMD simulations and measurements across a broad temperature range underlines this method's potential to predict elastic properties accurately. The strong anharmonicity implied by $\alpha(T)$ at extreme $P,T$s demands the application of SCAN-DPMD for such calculations.

\subsection{Elastic tensor}

\begin{figure*}[htbp]
    \centering
    \includegraphics[width=.8\textwidth]{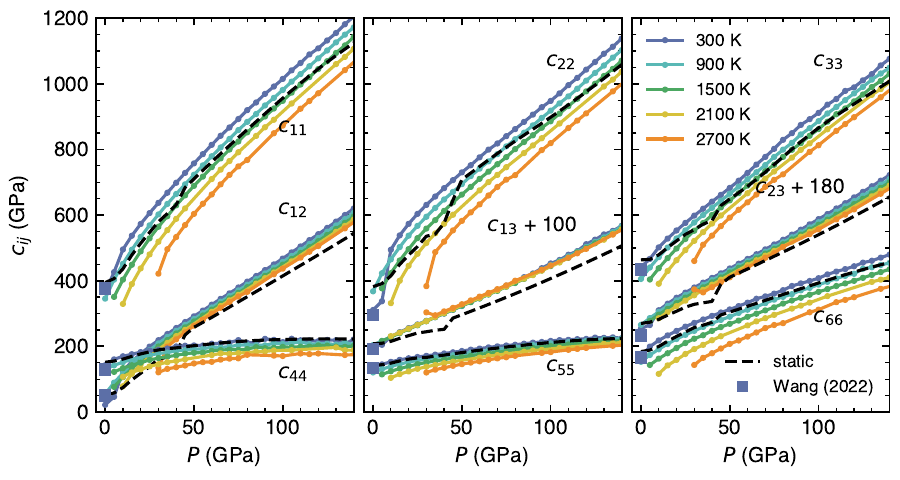}
    \caption{Comparison between static elastic tensor (black dashed curves) and adiabatic thermoelastic tensor components (solid-colored curves) of $\delta$-AlOOH calculated using SCAN-DP. Ultrasound measurements at ambient conditions on single crystal reported on \citep{wangSinglecrystalElasticityPhase2022} are included for reference.}
    \label{fig:elasticity}
\end{figure*}

$\delta$-AlOOH is an orthorhombic crystal, characterized by an elastic tensor ($C_{ij}$) with nine distinct components: $c_{11}$, $c_{22}$, $c_{33}$, $c_{12}$, $c_{12}$, $c_{23}$, $c_{44}$, $c_{55}$,  and $c_{66}$ \cite{bruggerPureModesElastic1965}. We evaluate these adiabatic tensor components using the stress-fluctuation method in DPMD simulations within the $NVE$ ensemble \citep{zhenDeformationFluctuationHybrid2012, clavierComputationThermalElastic2023, rayStatisticalEnsemblesMolecular1984, rayMolecularDynamicsCalculation1985, rayCalculationElasticConstants1986}. For compactness, we use the Voigt notation ($C_{ij}$, $i,j = 1~\mathrm{to}~6$) instead of the full cartesian notation ($C_{ijkl}$, $i,j,k,l=1,2,3$).
The elastic tensor presented in this section was the adiabatic effective elastic tensors ($C^S_{ij}$) derived using the stress-fluctuation method. Our comparisons indicate that all these methods give similar results. For a detailed discussion and comparison of these methods, please refer to Sections \ref{sec:stress-strain-result} and \ref{sec:adiabatic-correction-results}.

\begin{table*}
\caption{Comparison of elastic constants and other properties from various studies.}
\label{tab:elast-ambient}
\begin{ruledtabular}
\begin{tabular}{lcccccc}
& Wang (2022) \citep{wangSinglecrystalElasticityPhase2022} & Mashino (2016) \citep{mashinoSoundVelocitiesDAlOOH2016} & Li (2006) \citep{liElasticOpticalProperties2006} & Tsuchiya (2009) \citep{tsuchiyaElasticPropertiesDAlOOH2009a} & This study & This study \\
& Single crystal & Polycrystal & GGA static & GGA static & SCAN static & SCAN 300~K MD \\
\hline
$\rho$~(g/cm$^3$) & 3.536(1) & 3.593 & 3.532 & 3.383 & 3.575 & 3.545 \\
\hline
$c_{11}$~(GPa) & 375.9(9) & - & 416 & 314 & 394.2 & 364.9 \\
$c_{22}$~(GPa) & 295.4(11) & - & 509 & 306 & 381.8 & 310.7 \\
$c_{33}$~(GPa) & 433.5(12) & - & 418 & 391 & 463.5 & 441.4 \\
$c_{44}$~(GPa) & 129.2(6) & - & 133 & 117 & 151.5 & 142.4 \\
$c_{55}$~(GPa) & 133.4(7) & - & 124 & 115 & 141.9 & 135.0 \\
$c_{66}$~(GPa) & 166.4(6) & - & 229 & 152 & 184.0 & 168.0 \\
$c_{12}$~(GPa) & 49.7(9) & - & 137 & 34 & 50.4 & 21.8 \\
$c_{13}$~(GPa) & 91.9(15) & - & 93 & 95 & 108.0 & 101.5 \\
$c_{23}$~(GPa) & 52.8(21) & - & 84 & 67 & 89.8 & 74.3 \\
\hline
$K_{\mathrm{V}}$~(GPa) & 166.0(13) & - & 219.1 & 155.9 & 191.1 & 168.0 \\
$G_{\mathrm{V}}$~(GPa) & 146.5(3) & - & 166.0 & 131.1 & 160.6 & 150.4 \\
$K_{\mathrm{R}}$~(GPa) & 159.8(48) & - & 216.4 & 128.8 & 189.2 & 159.7 \\
$G_{\mathrm{R}}$~(GPa) & 144.0(15) & - & 157.4 & 123.0 & 159.7 & 148.0 \\
$K_{\mathrm{VRH}}$~(GPa) & 162.9(31) & - & 217.7 & 153.5 & 191.1 & 163.8 \\
$G_{\mathrm{VRH}}$~(GPa) & 145.2(13) & - & 161.7 & 130.0 & 160.6 & 149.2 \\
\hline
$V_P$~(km/s) & 10.04(7) & 9.54(7) & 11.11 & 9.83 & 10.65 & 10.11 \\
$V_S$~(km/s) & 6.41(3) & 5.89(10) & 6.77 & 6.20 & 6.70 & 6.49 \\
\end{tabular}
\end{ruledtabular}
\end{table*}

In Table~\ref{tab:elast-ambient}, we compare the computed $C_{ij}$ components at 0~GPa under static and 300~K conditions to previous DFT-GGA/PBE calculations \citep{tsuchiyaElasticPropertiesDAlOOH2009a, liElasticOpticalProperties2006} and ultrasound single-crystal measurements \cite{wangSinglecrystalElasticityPhase2022}.
Tsuchiya~\emph{et al.}'s GGA/PBE-based predictions \citep{tsuchiyaElasticPropertiesDAlOOH2009a} significantly underestimate the elastic stiffness tensor component by 15--30\%, and in some cases, by as much as 60~GPa, compared to single-crystal measurements \cite{wangSinglecrystalElasticityPhase2022} and our SCAN predictions. Their underestimation could be attributed to the PBE's overestimation of pressure, $\sim$5--6~GPa  \citep{luoInitioInvestigationHbond2022}, producing a larger equilibrium volume, $V_0$, and smaller bulk modulus, $K_0$. Conversely, Li~\emph{et al.}'s GGA/PBE-based results \citep{liElasticOpticalProperties2006} are generally more significant, particularly for the components $c_{11}$, $c_{22}$, and $c_{33}$. The reasons for this discrepancy between similar calculations are unclear, but these studies used different computational tools and pseudopotentials: Li~\emph{et al.}~\citep{liElasticOpticalProperties2006} employed VASP with the Projector Augmented Wave (PAW) pseudopotentials, whereas Tsuchiya~\emph{et al.}~\citep{tsuchiyaElasticPropertiesDAlOOH2009a} utilized \textsc{Quantum ESPRESSO} with norm-conserving pseudopotentials. Additionally, using different H-bond configurations that coexist at 0~GPa \citep{tsuchiyaVibrationalPropertiesDAlOOH2008, luoInitioInvestigationHbond2022} could contribute to the observed differences.

Benefitting from SCAN's more faithful reproduction of the compression curve, we have achieved a more accurate prediction of $V_0$.
Our fully anharmonic predictions of $C^S_{ij}$ at 300~K resemble much more closely the single-crystal ultrasound measurements \citep{wangSinglecrystalElasticityPhase2022}. Differences between our predictions and these measurements, $\sim$10--20~GPa, are all within the experimental uncertainties, except for $c_{22}$, which is $\sim$10 GPa outside the experimental uncertainty range. Overall, the agreement is excellent, especially considering that the joint experimental determination of these coefficients can compensate for underestimating some coefficients by overestimating others.

Fig.~\ref{fig:elasticity} shows the pressure-dependence of $\delta$'s adiabatic elastic tensor $C^S_{ij}$ at various temperatures. At 300~K, within the 0--15~GPa pressure range, the diagonal components $c_{11}$ and $c_{22}$ start softer than $c_{33}$ but undergo a steep increase under pressure. This pattern underpins the axial compression trends observed in Fig.~\ref{fig:1}(b) and in experiments \citep{sano-furukawaChangeCompressibilityDAlOOH2009}, where the $a$ and $b$ axes are notably more compressible within the same pressure range. Conversely, $c_{33}$, off-diagonal, and shear components exhibit milder anomalies under pressure, indicating they are less affected by the disorder-symmetrization transition. The higher compressibility pressure range extends to $\sim$15~GPa, $\sim$7--9~GPa beyond the experimental transition pressure at $\sim$6--8~GPa at 300~K \citep{sano-furukawaDirectObservationSymmetrization2018, sano-furukawaChangeCompressibilityDAlOOH2009}).

Except for $c_{44}$ and $c_{55}$, which nearly plateau at higher pressures, $C_{ij}$'s become more linearly dependent on pressure above 15~GPa at 300~K. Here, $c_{11}$ and $c_{22}$ become similarly stiff or stiffer than $c_{33}$. At around $\sim$90~GPa, we observed a subtle change in the rate of stiffening in $c_{22}$ and $c_{33}$: first, $P < 90$~GPa is a deceleration regime, then, after $P > 90$~GPa is an acceleration regime. They are likely associated more with the system's anharmonicity until $\delta$'s ``full symmetrization'', i.e., the potential energy surface for proton at the center of two O-ions becomes truly harmonic at $\sim$100~GPa \citep{luoProbingStateHydrogen2024}. It affects $\delta$'s $c_{22}$ and $c_{33}$ rate of increase with pressure but does not change the overall monotonic trend.

Between 900--1,500~K, the observed anomaly and subtle behavior of the
$C_{ij}$ tensor become less pronounced, yet they persist. Above 2,100~K and below 40~GPa, the steep increases in $C_{ij}$ intensify across all components as a precursor to the instabilities and potential dissociation or dehydration observed experimentally in this $P,T$ range \citep{sanoAluminousHydrousMineral2008, duanPhaseStabilityThermal2018, pietDehydrationDAlOOHEarth2020}.

Compared to the 300~K SCAN-DPMD simulations, static SCAN-DP underestimates $C_{ij}$ components across the entire pressure range. This underestimation is due to the missing H-bond disorder and dynamic effects \citep{sano-furukawaDirectObservationSymmetrization2018}, which leads to an overestimation of the transition pressure by $\sim$30~GPa. The discrepancy reinforces the necessity of employing MD to capture finite-temperature effects, especially across a phase transition, for accurate modeling. Notably, our static SCAN-DP's prediction is qualitatively similar to previous static GGA/PBE-DFT studies, which also report the jump in $C_{ij}$'s around 30~GPa \citep{tsuchiyaElasticPropertiesDAlOOH2009a}.

\subsection{Elastic moduli, acoustic velocities, and anisotropies}

\begin{figure*}[htbp]
    \centering
    \includegraphics[width=.8\textwidth]{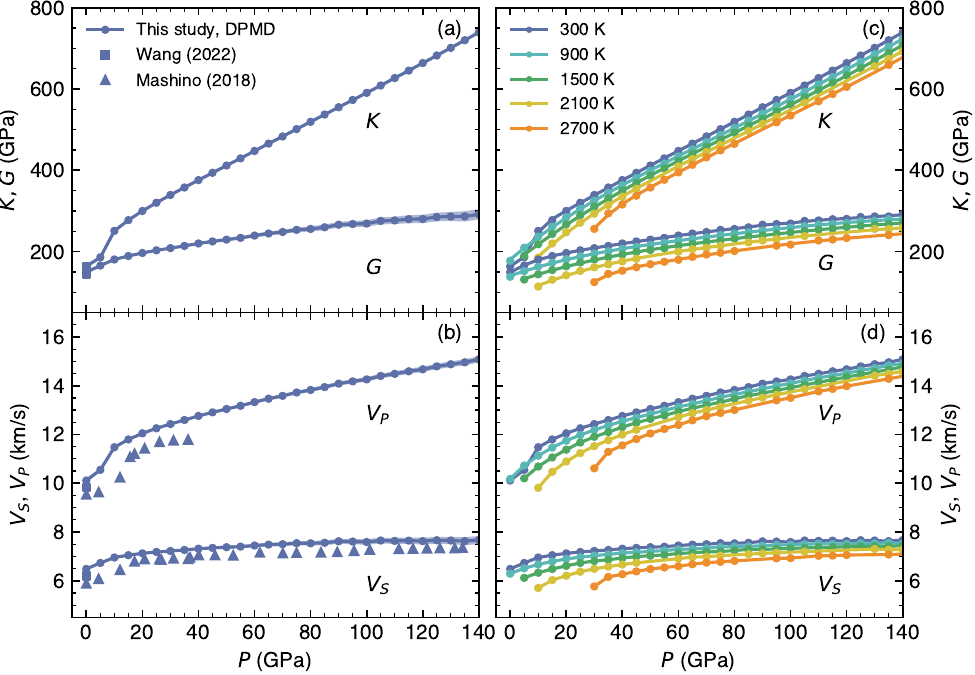}
    \caption{Predicted (a) bulk modulus ($K$) and shear modulus ($G$), (b) compressional velocity ($V_P$), and shear velocity ($V_S$) of $\delta$-AlOOH vs.\ $P$ predicted by SCAN-DPMD at 300~K compared to ultrasound measurements on a single-crystal sample at ambient conditions by \citet{wangSinglecrystalElasticityPhase2022} and Brillouin scattering under pressure on a polycrystalline sample at room temperature by \citet{mashinoSoundVelocitiesDAlOOH2016}.  (c, d) are similar predictions at high temperatures. The shaded regions indicate the Voigt-Reuss-Hill (VRH) bounds at 300~K.}
    \label{fig:elastic-moduli-and-velocity}
\end{figure*}

Based on the calculated $C^S_{ij}$, we first determine the bulk and shear moduli ($K$ and $G$) using Eqs.~\eqref{eq:hill-elastic-moduli}--\eqref{eq:voigt-reuss-elastic-moduli}, then determine the compressional and shear velocities ($V_P$ and $V_S$) using Eq.~\eqref{eq:velocity}. Fig.~\ref{fig:elastic-moduli-and-velocity} shows these results vs.\ $P,T$.

At 300~K within the 5--10~GPa range, we observe a pronounced ``jump'' in both $K$ and $V_P$, alongside a more gradual yet anomalous increase in $G$ and $V_S$. These can be attributed to the anomalous changes in $C^S_{ijkl}$ and $P(V)$ compression curve in this pressure range as discussed above (see Fig.~\ref{fig:1}). 
Beyond 30~GPa, all properties, $K$, $G$, $V_S$, and $V_P$, display a smooth monotonic increase with pressure. 

The elastic moduli and acoustic velocities all soften with increasing temperature. 
However, we do not identify a signal related to the onset or enhancement of the protons' superionic behavior \citep{luoProbingStateHydrogen2024}.
This is confirmed by fitting $V_P$ and $V_S$ above 90~GPa to a polynomial second-order in $P$ and first-order in $T$, $f(P, P^2, T)$. Without a super-linear $T$ dependence, the model still describes $V_P$ and $V_S$ at high $P,T$s quite accurately, even above the superionic transition boundary starting at $\sim$2,100~K at $\sim$60~GPa \citep{luoProbingStateHydrogen2024}. The model parameters for $V_P$ and $V_S$ are given in Table~SI, and their validation is shown in Fig.~S1. The term linear in $T$ captures the $V_S$'s and $V_P$'s dependences on $T$ over the entire pressure range quite well. 
The deviation from the linear model observed at $V_P < 12$~km/s and $V_S < 7$~km/s corresponds to $P < 15$~GPa at 300~K and up to $P < 45$~GPa at 2,700~K, a regime where $\delta$'s H-bonds are asymmetric and $V_P$ and $V_S$ behave differently from higher pressures where H-bonds are symmetric.
The absence of higher-order dependence on $T$ suggests that the Al-O framework predominantly dictates the overall properties of $\delta$ after the H-bond disorder-symmetrization transition. This observation explains why $\delta$'s elastic properties are more similar to those of typical non-hydrous high-pressure systems rather than those of hydrous phases or NAMs \citep{duanPhaseStabilityThermal2018, mashinoSoundVelocitiesDAlOOH2016}.

Compared to measurements, our predicted $V_P$ and $V_S$ align closely with recent ambient-condition single-crystal measurements \citep{wangSinglecrystalElasticityPhase2022} but are faster than those obtained from polycrystalline measurements \citep{mashinoSoundVelocitiesDAlOOH2016}.
Our results systematically overestimate these parameters by 5\% across the entire pressure range of measurements on the polycrystalline sample. This is not unexpected since grain boundaries and pores are known to reduce acoustic velocities \citep{gleasonAnomalousSoundVelocities2011, marquardtElasticPropertiesMgO2011, wangSinglecrystalElasticityPhase2022}. The Voight-Reuss bounds outlined in shaded color appear marginal for this system and cannot accommodate such discrepancies.

\begin{figure}[htbp]
    \centering
    \includegraphics[width=.40\textwidth]{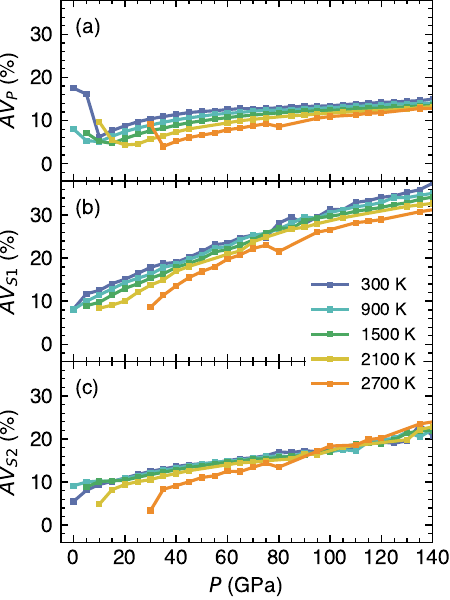}
    \caption{$P,T$ dependencies of $\delta$-AlOOH's azimuthal anisotropy factors ($AV$s) for $\delta$-AlOOH vs.\ $P$ predicted by SCAN-DPMD simulations.
    }
    \label{fig:anisotropy}
\end{figure}

The azimuthal anisotropy of each acoustic velocity mode is further quantified by the anisotropy factor ($AV$), defined by
\begin{equation}
    AV = \frac{V_\mathrm{max} - V_\mathrm{min}}{V_\mathrm{max} + V_\mathrm{min}} \times 200\% \,,
\end{equation}
where $V_\mathrm{max}$ and $V_\mathrm{min}$ denote the maximum and minimum velocities over all azimuthal directions. The three velocity modes, $V_P$, $V_{S1}$, and $V_{S2}$, are the solutions to the Christoffel's equation \citep{musgraveCrystalAcousticsIntroduction1970} at direction represented by the unitary vector $\hat{n} = (n_1, n_2, n_3)$,
\begin{equation}
\left| C_{ijkl} n_j n_l - \rho V^2 \delta_{ik} \right| = 0 \,.
\end{equation}
Fig.~\ref{fig:anisotropy} shows $AV_P$, $AV_{S1}$, and $AV_{S2}$ vs.\ $P$. The $AV$s generally increase with pressure except for dips in $AV_P$ and a jump in $AV_{S2}$ in the 0--40~GPa range, depending on the temperature. The dips correspond to the anomalies in $C^S_{ij}$, but their effects are enhanced here. The necking in $C_{ij}$'s seen in Fig.~\ref{fig:anisotropy} likely corresponds to the end of the complete disorder-symmetrization transition, which changes the pressure dependence of $AV_P$ from super-linear to linear. The $AV$s slightly decrease with increasing temperature. 
$\delta$ is significantly less anisotropic than serpentines \cite{dengElasticAnisotropyLizardite2022, mookherjeeTrenchParallelAnisotropy2011} but considerably more anisotropic than the major anhydrous mantle components, e.g., MgO periclase and \cite{karkiHighpressureElasticProperties2001} and MgSiO$_3$ perovskite \citep{wanThermoelasticPropertiesBridgmanite2024, karkiHighpressureElasticProperties2001}.

\subsection{The stress vs.\ strain method}
\label{sec:stress-strain-result}

We compare the adiabatic tensor, $C^S_{ijkl}$, derived using the stress-strain relation with the one derived from the stress-fluctuation method. We expect that the result might differ slightly from the stress-fluctuation method because energy is not fully conserved during the simulation box deformation.
The first law of thermodynamics states that
\[
\Delta U = Q - W = T \Delta S - P \Delta V ~,
\]
where $\Delta U$ represents the change in the internal energy, $U = E_\mathrm{k} + E_\mathrm{p}$, and $W = -P \, \Delta V$ is the work done by the system.
A true isentropic process implies $\Delta S = 0$, and $\Delta U = W = -P \, \Delta V$. Both $\Delta U$ and $W = -P \, \Delta V$ can be obtained throughout the deformation simulation, and our calculation indicates that the condition $\Delta U = -P\,\Delta V$ is not strictly guaranteed, but it is close (see Fig.~S2). To ensure a proper isentropic process, adjustments to $E_\mathrm{k}$ (the scaling of the velocities) are necessary. This implies potential discrepancies with the stress-fluctuation method, even when running a MD simulation in a constant energy ($NVE$) ensemble.

\begin{figure*}
    \centering
    \includegraphics[width=.85\textwidth]{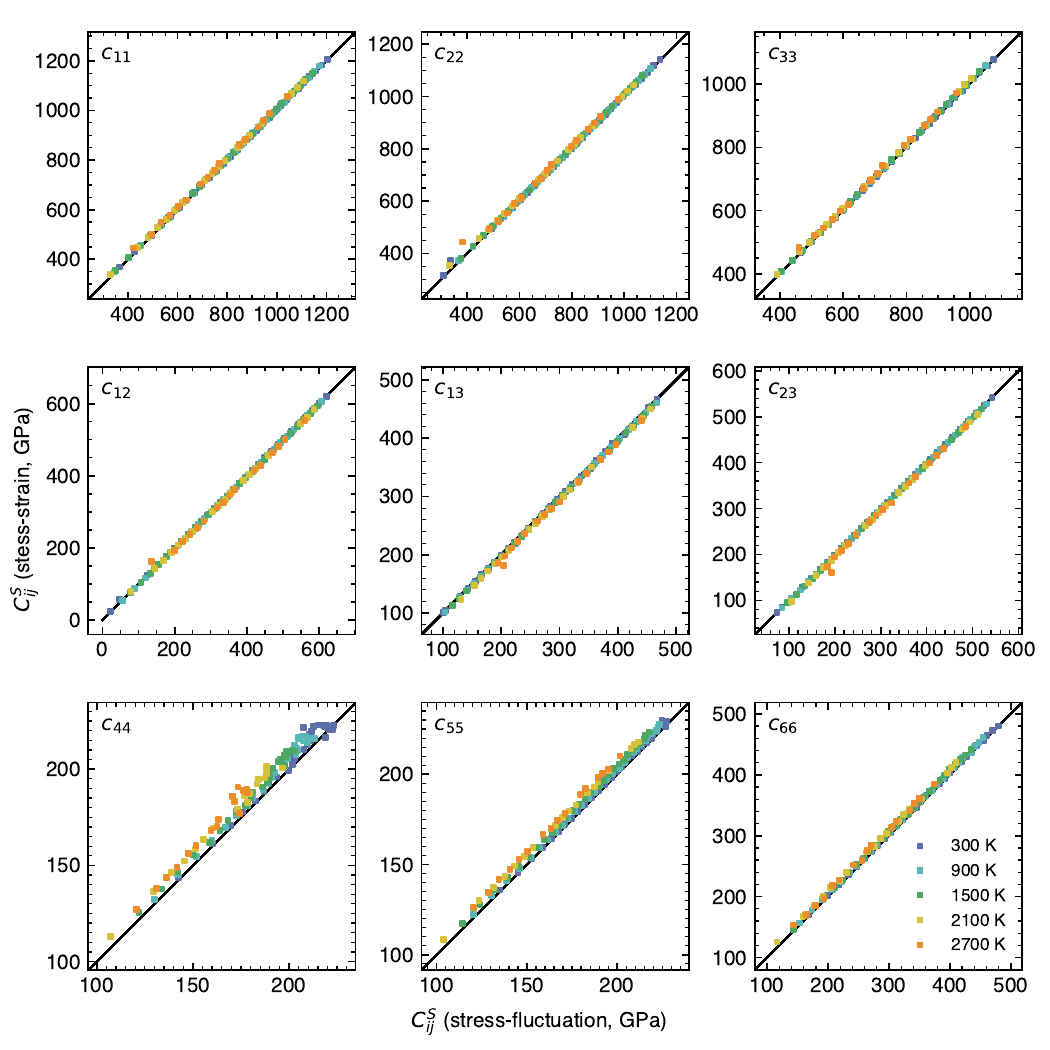}
    \caption{Comparison of $C^S_{ij}$ components obtained with stress-fluctuation and the stress-strain approaches using SCAN-DPMD within the $NVE$ ensemble at various $P,T$ conditions.}
    \label{fig:strain-vs-fluctuation}
\end{figure*}

Compared to the stress-fluctuation method (Fig.~\ref{fig:strain-vs-fluctuation}), we observe a deviation of less than 10~GPa in $C^S_{ij}$ components. This deviation is quite small, or less than 2\%, for diagonal and off-diagonal components. Still, it translates to a 10\% overestimation for the shear components due to their relatively small absolute values. The horizontal spread indicates the challenge in converging the stress-fluctuation calculations for the $c_{44}$ and $c_{55}$ components (see Fig.~S3).

Practically, the stress-fluctuation method offers a streamlined approach for obtaining $C^S_{ijkl}$ through MD simulations, eliminating the need for multiple MD simulations and offering a mathematically rigorous solution. Our tests suggest that employing both a larger simulation cell and an extended simulation timescale enhances the convergence of stress-fluctuation calculations. The Born matrix term, i.e., $\langle A^B_{ijkl} \rangle$ in Eq.~\eqref{eq:stress-fluctuation}, converges rapidly to an uncertainty $< 0.1$~GPa, often within a few thousand timesteps (e.g., Fig.~S3). However, converging the stress-fluctuation term (i.e., $V / k_B T \big[ \langle \sigma_{ij} \, \sigma_{kl} \rangle - \langle\sigma_{ij}\rangle\,\langle\sigma_{kl} \rangle \big]$) proves to be substantially more challenging, particularly at low-$T$ and high-$P$ conditions, where low-frequency vibrational modes are relatively more important. Small timesteps must be used in systems such as $\delta$, which contain light and faster-moving ions,  to ensure numerical accuracy in the equation of motion integration. The simultaneous need for short timesteps (0.2--0.5~fs) and long simulation run times ($\sim$10~ns) to converge the stress covariance term in Eq.~\eqref{eq:stress-fluctuation} make these simulations more challenging (see convergency test in Fig.~S3). In this case, the less straightforward stress vs.\ strain approach to compute $C^S_{ij}$ is useful.

\subsection{Adiabatic vs.\ isothermal elastic tensor}
\label{sec:adiabatic-correction-results}

\begin{figure*}
    \centering
    \includegraphics[width=.85\textwidth]{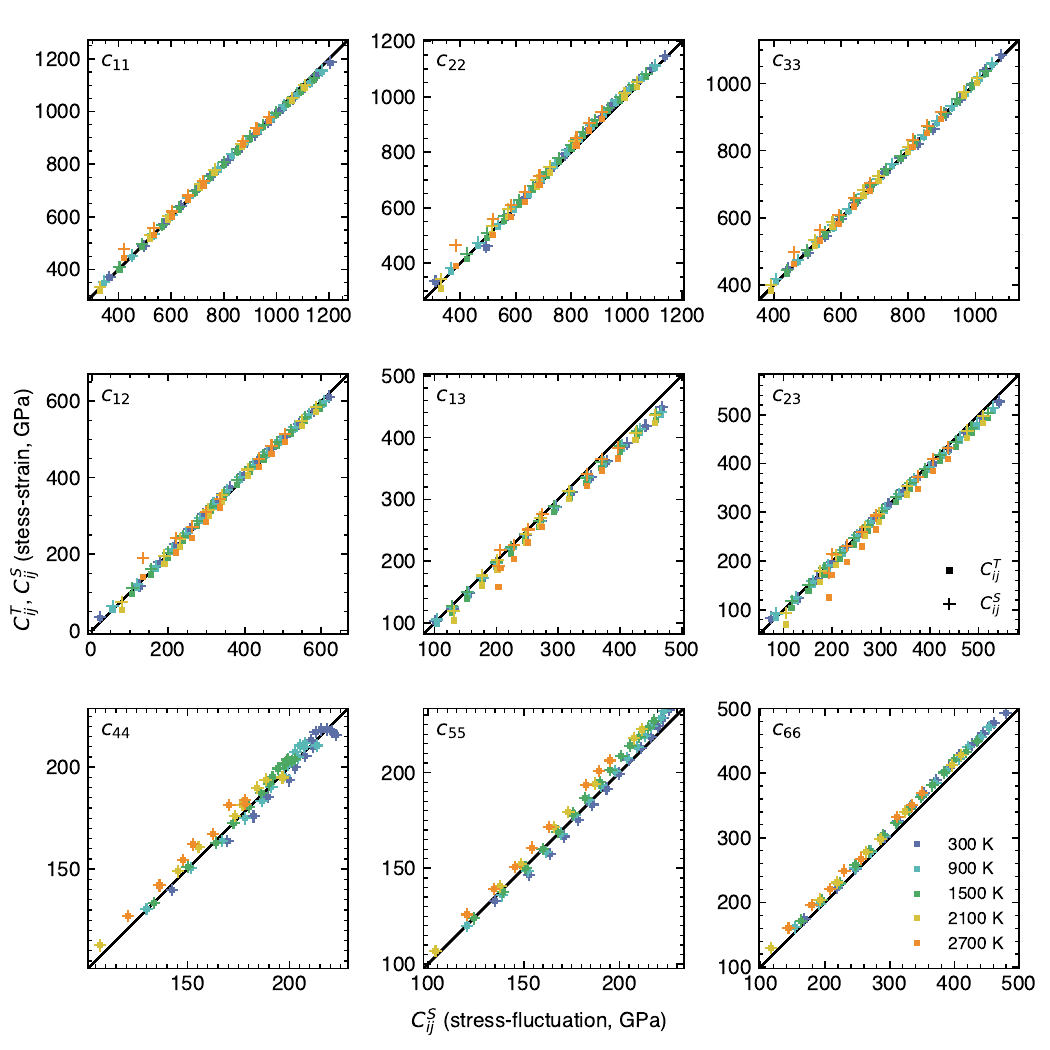}
    \caption{Comparison between adiabatic elastic tensor components, $C_{ij}^S$, computed in two different ways: a)~derived from stress fluctuations within the $NVE$ ensemble ($x$-axis), and b)~computed from the isothermal tensor, $C_{ij}^T$, computed from stress-strain relations in the $NVT$ ensemble followed by the adiabatic correction, i.e., the second term in Eq.~\eqref{eq:adiabatic-correction} ($y$-axis). Squares show $C_{ij}^T$- and crosses show $C_{ij}^S$-components.}
    \label{fig:adiabatic-vs-isothermal}
\end{figure*}

To compare the adiabatic and isothermal elastic tensors and to test the adiabatic correction, we also performed stress-strain calculations within the $NVT$ ensemble to derive the isothermal tensor, $C^T_{ijkl}$, then applied the adiabatic correction [Eq.~\eqref{eq:adiabatic-correction}] to derive $C^S_{ijkl}$.
We compared these results to $C^S_{ijkl}$ derived from the stress-fluctuation analysis in Fig.~\ref{fig:adiabatic-vs-isothermal}.
For diagonal components, $C^T_{ijkl}$ obtained from the strain-stress relation does not significantly deviate from $C^S_{ijkl}$. The second term in Eq.~\eqref{eq:adiabatic-correction} increases their values slightly, but the overall change is not significant. Results at low $P$ and high $T$ become even worse compared to $C^S_{ijkl}$ obtained using the stress-fluctuation analysis. The effect is more pronounced at higher $T$s when thermal expansivity increases anomalously with hydrogen diffusion (see Fig.~\ref{fig:alpha-temp}) \citep{luoProbingStateHydrogen2024}.
For the off-diagonal components ($c_{12}$, $c_{13}$, and $c_{23}$), the second term in Eq.~\eqref{eq:adiabatic-correction} noticeably mitigates temperature-dependent discrepancies. It significantly improves the agreement with $C^S_{ijkl}$ obtained from the stress-fluctuation analysis.
The correction does not affect shear components. They behave similarly to $C^S_{ijkl}$ obtained via the stress-strain relation directly in the $NVE$ ensemble (Fig.~\ref{fig:strain-vs-fluctuation}).
Overall, the adiabatic correction in Eq.~\eqref{eq:adiabatic-correction} is necessary at high temperatures, especially for superionic systems.
Practically, deriving $\alpha_i$ numerically presents significant challenges. Due to the absence of an explicit relation, deriving $\alpha_i$ from the box shapes requires dense $P,T$ sampling, particularly for systems at high temperatures where $\alpha_i$ exhibits a super-linear temperature dependence $T$. In comparison, the stress-fluctuation calculations are more straightforward and more manageable.

\section{Conclusion}
\label{sec:conclusion}

This study employs SCAN-DPMD simulations to investigate the thermoelastic properties and acoustic velocities of $\delta$-AlOOH at the $P,T$s up to 2,700~K and 145~GPa, corresponding closely to subducting slab conditions in the mantle. The predicted $C_{ij}$ components agree well with previous single-crystal ultrasound measurements at ambient conditions \citep{wangSinglecrystalElasticityPhase2022}. The pressure dependence of sound velocities at high pressures is similar to those measured at 300~K in polycrystalline samples by Brillouin scattering \citep{mashinoSoundVelocitiesDAlOOH2016}, including the steep increase in $K$ and $V_P$ caused by the disorder-symmetrization phase transition at $\sim$10~GPa and 300~K. However, the predicted velocities are faster than the Brillouin scattering measurements. by $\sim$5\%. The difference falls outside the Voigt-Reuss bounds and likely originates in significant effects caused by grain size, grain boundaries, and possibly pores. The impact of proton diffusion at high temperatures on sound velocities does not seem obvious.

Under $P,T$s relevant to the Earth's lower mantle, $\delta$ has a significantly faster velocity than typical hydrous minerals. Its elastic properties are more similar to those of anhydrous phases, e.g., stishovite-type SiO$_2$. The phase equilibrium between phase Egg (AlSiO$_3$OH) \citep{schmidtSynthesisCrystalStructure1998, onoHighTemperatureStability1999, vanpeteghemCompressibilityPhaseEgg2003a, xueCationOrderHydrogen2006}, $\delta$-AlOOH, and stishovite-type SiO$_2$ with hydrous defects \citep{spektorUltrahydrousStishoviteHighpressure2011, spektorFormationHydrousStishovite2016, nisrPhaseTransitionEquation2017, nisrRamanSpectroscopyWaterrich2017, zhangElasticityHydrousSiO22024, palfeyBehaviorHydrogarnetType2023}, i.e., $\mathrm{Egg} \rightleftharpoons \delta\hbox{-AlOOH} + \mathrm{Stv}$, is relevant to the Earth's deep water cycle \citep{ishiiSuperhydrousAluminousSilica2022, wangSinglecrystalElasticityPhase2022} and it is desirable to recognize it in seismic tomography. Because of phase Egg's significantly slower sound velocity \citep{mookherjeeAnomalousElasticBehavior2019, wangSinglecrystalElasticityPhase2022} than those of $\delta$ and stishovite, the formation (or decomposition) of phase Egg in the mantle would cause a decrease (or increase) in velocity. Since this behavior is different from near ambient conditions, and single crystal elasticity measurement is only available at ambient pressure and temperatures before the pressure-induced H-bond transitions in $\delta$, single crystal measurements at high pressures and temperatures after the phase transitions are highly desirable.

MD simulations are necessary to investigate systems with strong anharmonicity at high temperatures. Due to its rigorous mathematical framework, the stress-fluctuation method with the $NVE$ ensemble remains a cornerstone among various MD methods for computing adiabatic thermoelastic properties. However, despite its trade-offs in energy conservation during deformation, the stress vs.\ strain approach within the $NVE$ ensemble still stands as a practical alternative, especially when convergence in stress fluctuation becomes more challenging.
Determining the isothermal thermoelastic tensor first via the strain-stress relations with the $NVT$ ensemble, followed by an adiabatic correction, is less practical because it involves calculations of linear thermal expansion coefficients, which, due to the lack of an explicit volume-temperature relationship, relies on numerical differentiation on a densely-sampled $P,T$ grid.
The differences in individual $C_{ij}$ components calculated by various methods are at most 20~GPa at all pressures, hardly affecting polycrystalline averages for elastic moduli and sound velocities.

\section*{Acknowledgments}

DOE Award DE-SC0019759 supported this work. Calculations were performed on the Extreme Science and Engineering Discovery Environment (XSEDE) \citep{townsXSEDEAcceleratingScientific2014} supported by the NSF grant \#1548562 and Advanced Cyberinfrastructure Coordination Ecosystem: Services \& Support (ACCESS) program, which is supported by NSF grants \#2138259, \#2138286, \#2138307, \#2137603, and \#2138296 through allocation TG-DMR180081. Specifically, it used the \textit{Bridges-2} system at the Pittsburgh Supercomputing Center (PSC), the \textit{Anvil} system at Purdue University, the \textit{Expanse} system at San Diego Supercomputing Center (SDSC), and the \textit{Delta} system at National Center for Supercomputing Applications (NCSA). We gratefully acknowledge Dr.\ Germain Clavier for the helpful discussion on thermoelasticity.

\bibliography{Geophysics}

\end{document}

\title{\bfseries\textsc{Supplementary Information}\vspace{16pt}\\
Elasticity and acoustic velocities of $\delta$-AlOOH\\
at extreme conditions: a methodology assessment
}

% \title{
% Elasticity and acoustic velocities of $\delta$-AlOOH\\
% at mantle pressures and temperatures
% }

\author{Chenxing Luo\,\orcidlink{0000-0003-4116-6851}}
\affiliation{Department of Applied Physics and Applied Mathematics, Columbia University, New York, New York 10027, USA}

\author{Yang Sun\,\orcidlink{0000-0002-4344-2920}}
\email[]{yangsun@xmu.edu.cn}
\affiliation{Department of Applied Physics and Applied Mathematics, Columbia University, New York, New York 10027, USA}
\affiliation{Department of Physics, Xiamen University, Xiamen, 361005, China}
% \affiliation{Department of Physics and Astronomy, Iowa State University, Ames, Iowa 50011, USA}

\author{Renata M.\ Wentzcovitch\,\orcidlink{0000-0001-5663-9426}}
\email[]{rmw2150@columbia.edu}
\affiliation{Department of Applied Physics and Applied Mathematics, Columbia University, New York, New York 10027, USA}
\affiliation{Department of Earth and Environmental Sciences, Columbia University, New York, New York 10027, USA}
\affiliation{Lamont--Doherty Earth Observatory, Columbia University, Palisades, New York 10964, USA}
\affiliation{Data Science Institute, Columbia University, New York, New York 10027, USA}
\affiliation{Center for Computational Quantum Physics, Flatiron Institute, New York, NY 10010, USA}

\maketitle
\thispagestyle{empty}

% \vfill

\textbf{This PDF file includes:}

\begin{itemize}
\item Tables SI
\item Figures S1 to S3
\end{itemize}

% \hbox{}

\newpage

\begin{table}[p]
    \caption{Parameter describing the SCAN-DPMD predicted sound velocities of $\delta$ fitted to $M_0 + (\partial M / \partial P) \, P + (\partial^2 M / \partial P^2) \, P^2 + (\partial M / \partial T) \, T$.}
    \label{tab:derivat}
    \begin{ruledtabular}
    \begin{tabular}{lcccc}
        % \toprule
        & $M_0$ (km/s)
        & $ \partial M / \partial P \times 10^{-2}$
        & $ \partial^2 M / \partial P^2 \times 10^{-5}$
        & $ \partial M / \partial T \times 10^{-4}$ \\
        & 
        & ($\mathrm{km \cdot s^{-1} \cdot GPa^{-1}}$)
        & ($\mathrm{km \cdot s^{-1} \cdot GPa^{-2}}$)
        & ($\mathrm{km \cdot s^{-1} \cdot K^{-1}}$) \\
        \midrule
        $V_P$ & $11.61$ & $3.28$ & $-5.06$ & $-3.10$ \\
        $V_S$ &  $6.85$ & $1.33$ & $-4.74$ & $-2.67$ \\
        % \bottomrule
    \end{tabular}
    \end{ruledtabular}
\end{table}

\newpage
\clearpage

\begin{figure}[p]
    \centering
    \includegraphics[width=.9\textwidth]{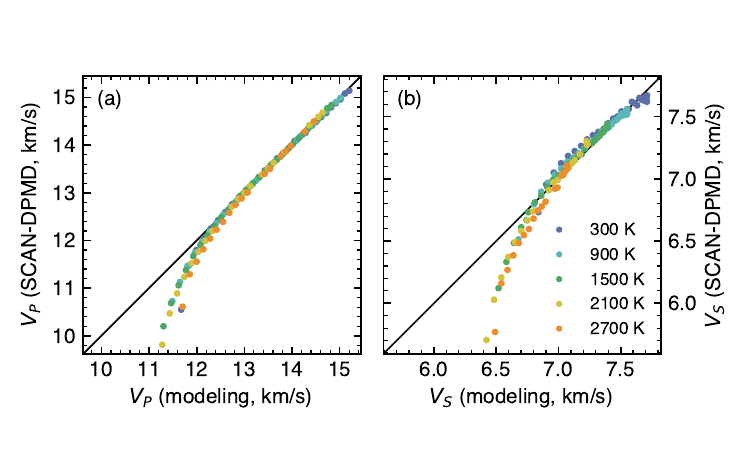}
    \caption{Modeling and validation of (a) $V_P$ and (b) $V_S$ at $T > 90$~GPa using a linear model of $P$, $P^2$, and $T$ and validation over the entire pressure range.}
    \label{fig:method}
\end{figure}

\newpage

\begin{figure}[p]
    \centering
    \includegraphics[width=.48\textwidth]{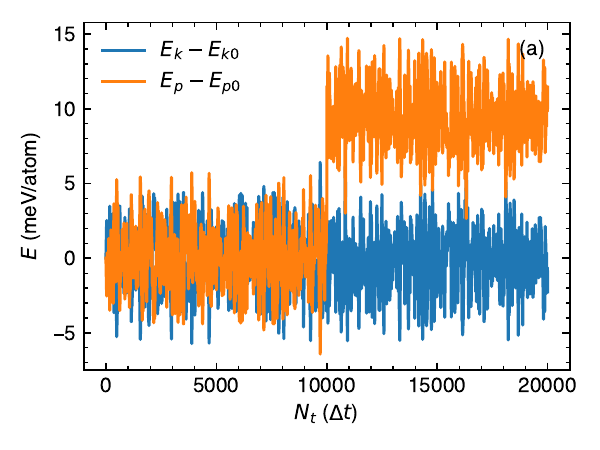}
    \includegraphics[width=.48\textwidth]{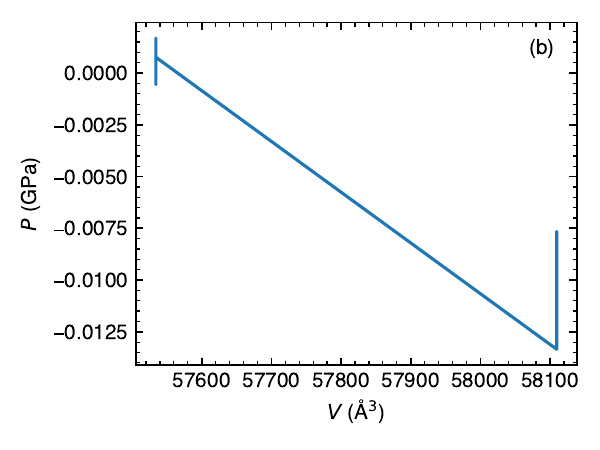}
    \includegraphics[width=.48\textwidth]{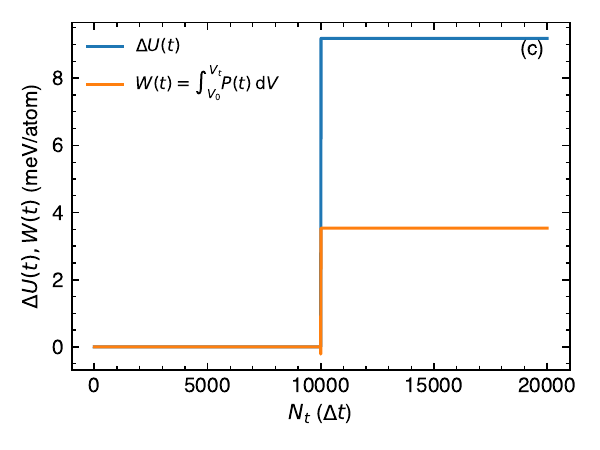}
    \caption{Analysis of the $NVE$ simulation deformation process with a 1,024-atom cell. For a system equilibrated with an $NVT$ simulation at 0~GPa and 300~K. We first performed a 10,000-step $NVE$ simulation. A 1\% strain was then applied to the system, after which another 10,000-step $NVE$ simulation was conducted using the deformed configuration. Here, we show (a) energy differences before and after deformation; (b) $P$-$V$ relation diagram throughout the deformation process; (c) calculation of the internal energy difference $\Delta U = \Delta E_k + \Delta E_p$ and the work done $W$ during deformation. The relationship $W = \Delta U$ should hold for an ideally adiabatic process.}
    \label{fig:method}
\end{figure}

\newpage

\begin{figure}[p]
    \centering
    \includegraphics[width=\textwidth]{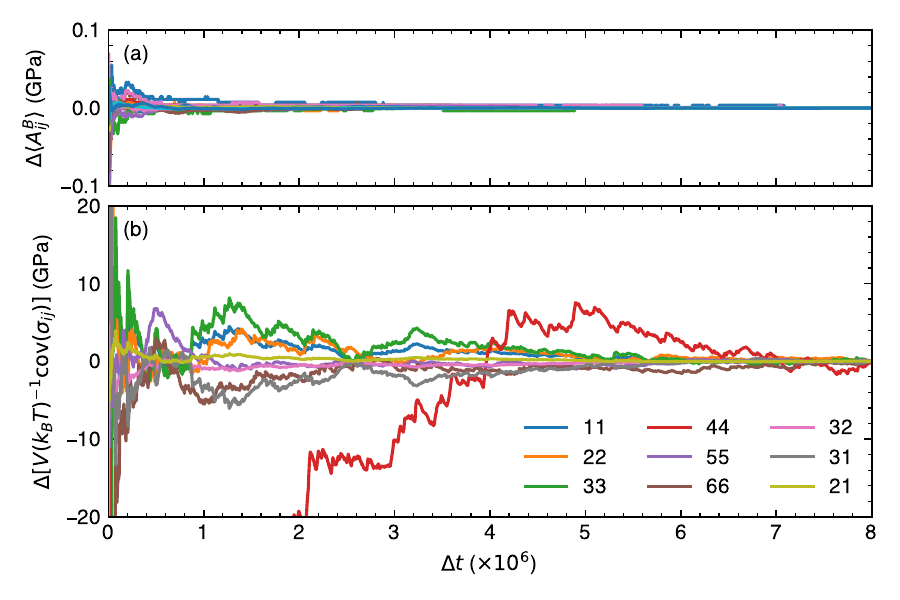}
    \caption{Convergence of (a) $\langle A^B_{ijkl} \rangle$ and (b) $V / k_B T \big[ \langle \sigma_{ij} \, \sigma_{kl} \rangle - \langle\sigma_{ij}\rangle\,\langle\sigma_{kl} \rangle \big]$ term components w.r.t.\ simulation time step (i.e., rolling averages compared to the final values) at 300~K, 105~GPa.}
    \label{fig:method}
\end{figure}

\clearpage

% \bibliography{Geophysics}